\documentstyle[preprint,floats,prl,aps]{revtex}

\input epsf.tex
\def\desepsf(#1 width #2){\epsfxsize=#2 \epsfbox{#1}}
\def \pom {{\scriptscriptstyle \kern -0.1em I \kern -0.25em P}}
\def\be{\begin{equation}}
\def\ee{\end{equation}}
\def\bea{\begin{eqnarray}}
\def\eea{\end{eqnarray}}
\begin{document}

\preprint{\vbox{
\hbox{PSU-TH/219} 
\hbox{October 1999}
\hbox{} }  }
\draft

\vskip 2 true cm
 
\title{Diffractive parton distributions  in light-cone 
QCD~\footnote{Talk by F. Hautmann at the International 
Symposium on  Multiparticle 
Dynamics and QCD ISMD99, Brown University, 9-13 August 1999, 
to appear in the proceedings.}}

\author{ F. Hautmann }

\address{Department of Physics, Pennsylvania State University,  
 University Park PA 16802}

\author{ Z. Kunszt }

\address{Institute of Theoretical Physics, ETH, 
CH-8093 Zurich}

\author{ D.E. Soper }

\address{Institute of Theoretical Science, University of Oregon, 
Eugene OR 97403}

\maketitle

\begin{abstract}
We discuss recent theoretical results on diffractive deeply 
inelastic scattering,   focusing on     the partonic picture of 
diffraction     in configuration space and the 
predictions for      
the $\beta$ behavior and the scaling violation.  
\end{abstract}  


\section{Introduction}

The understanding of 
hadronic diffraction within QCD is one of the frontiers 
of the theory of the strong interactions.  
A great deal of  progress  has been achieved recently in this area, 
following the 
realization that, for 
diffractive hard processes with only one 
hadron in the initial state, 
a  property of factorization 
of short and long distances holds, 
similar to the analogous property in 
inclusive hard scattering~\cite{halina}.  
This property allows one to give a parton interpretation 
of diffractive deeply inelastic scattering. 
This talk is based on work~\cite{hkslong} in which the 
methods of light-cone QCD are used to investigate diffractive 
parton distributions. 
 Sec.~2 
 is devoted to discussing  
  the basic elements of this approach. 
Secs.~3 and 4 describe applications to, respectively, the case of  
a gedanken diffraction experiment and the case of HERA collider 
experiments.

\section{Factorization and ``aligned jet'' picture  }

Consider diffractive deeply inelastic scattering of a hadron $A$,  
\begin{equation}
\label{ddis}
 e + A \to e^\prime + A^\prime + X, 
\end{equation}
in which 
$A$ is  scattered with a fractional loss of longitudinal 
momentum $x_\pom$  
and invariant momentum transfer $t$. Let   
$Q^2$ and $x$ be as usual the photon virtuality and the Bjorken variable 
of the deeply inelastic collision.   
The factorizability of the hard scattering~\cite{proo} implies that  
the diffractive deeply inelastic cross section can be written 
in terms of diffractive parton distributions, that is, 
functions that represent the probability to find a parton of type $a$ 
in hadron $A$  on the condition that $A$ is diffractively 
scattered~\cite{bere}.     

These distributions can be defined as matrix elements of certain 
``measurement" operators, given through  the fundamental quark 
and gluon  fields. 
If $p_A$ is the hadron's momentum, taken to be in the 
plus light-cone direction, and $\beta x_\pom$ is the fraction 
of this momentum carried by the parton, 
 the form of the  gluon distribution is  
\begin{eqnarray}
\label{opg}
&& {{d\, f_{g/A}^{\rm {diff}}(\beta x_\pom, x_\pom , t , \mu ) } \over
{dx_\pom\,dt}} =  
{1 \over (4\pi)^3 
\beta x_\pom p_{\!A}^+}\sum_{X} \int d y^-
e^{i\beta x_\pom p_{\!A}^+ y^-}
\\
&& \hskip 1 cm  \times 
\langle A |\widetilde G_a(0)^{+j}
| A^\prime, X \rangle 
\langle A^\prime, X|
\widetilde G_a(0,y^-,{\bf 0})^{+j}| A \rangle 
 ,  
\nonumber
\end{eqnarray}
where $\widetilde
G_a^{+j}$ is the field strength  modified by
multiplication by 
the path-ordered exponential of a line integral of the color potential:
\begin{equation}
\label{Gtilde}
\widetilde G_a(y)^{+j}
=
E(y)_{ab}
G_b(y)^{+j} \hspace*{0.2 cm} ,  \hspace*{0.2 cm} 
%%
%\end{equation}
%\begin{equation}
%%
%\label{eikdef}
E(y) = 
{\cal P}
\exp\left(
- i g \int_{y^-}^\infty d x^-\, A_c^+(y^+,x^-,{\bf y})\, t_c
\right) \hspace*{0.2 cm} .
\end{equation}
 The scale $\mu$ in Eq.~(\ref{opg}) is the 
  scale at which the ultraviolet 
  divergences from the operator products are renormalized.    
  A formula analogous to (\ref{opg}) 
 holds for the quark distribution. 

Eq.~(\ref{opg}) may be understood starting from 
the theory canonically quantized on planes $x^+ = {\mbox {const.}}$ 
in  $A^+ = 0$ gauge. In this gauge the operator in 
Eq.~(\ref{opg})  
can be related to the number operator. In general,  though, 
the definition in Eq.~(\ref{opg}) 
is  gauge-invariant: it is made gauge-invariant by 
the path-ordered exponential (\ref{Gtilde}). This factor  has a 
physical    interpretation  in terms of the  
recoil color flow~\cite{hkslong,bere},  
represented as    a 
fast moving color charge 
that goes out along a lightlike line in the minus direction,  
coupling to gluons from the hadron's field.

The usefulness of the definition (\ref{opg}) 
(and its  counterpart for quarks) 
comes from the fact 
that the   experimentally observed 
diffractive structure functions,  
$ d F^{\rm diff}_k / [ d x_\pom \, dt]$,  
are related to $df_{a/A}^{\rm {diff}}/ [dx_\pom\,dt]$ 
through short-distance coefficients 
${\hat F}$ which are  known from 
perturbative calculations for deeply inelastic 
scattering~\cite{neerven}:  
\begin{equation} 
\label{fact} 
 { { d F^{\rm diff}_k 
 } \over { dx_\pom\, dt}}
\sim  {\hat F}_{k a}  \otimes 
  { df_{a/A}^{\rm diff} 
\over dx_\pom\, dt} 
\hspace*{0.3 cm} . 
\end{equation}
 That is, the definition (\ref{opg}) projects out precisely the 
long-distance factor that distinguishes diffractive deeply inelastic 
scattering from ordinary (inclusive) deeply inelastic scattering.

To gain insight into the structure of this  factor, it is useful  
to look at   the diffraction process  in a
reference frame in which the struck hadron is at rest (or almost 
at rest). The aligned jet picture~\cite{alignedjet}  of 
deeply inelastic scattering  suggests that in such a frame the 
spacetime structure of the  small $x$ process looks simple.   
 The  parton 
probed by the measurement operator is created at light cone 
times $x^-$ far in the past, $x^- \to - \infty$, along with a 
color source of the opposite color. This system  
moves with large  momentum in the minus direction, of order
$ k^- \sim \langle p_\perp \rangle  / x _\pom  $, 
with $\langle p_\perp \rangle$ being the typical transverse momentum of 
the system. It   
 evolves slowly in $x^-$, possibly turning into a system with 
more partons. 
  Much later, around $x^- \approx 0$,  it interacts with the 
color field produced by the diffracted   hadron. 
After that, it continues its slow evolution. 

In the simplest  perturbative approximation to this picture, 
only one gluon is emitted 
into the final state, and the interaction with the color field 
of the hadron is simply given  by the absorption of two gluons. 
This case is depicted in Fig.~1.

\begin{figure}[htb]
\vspace{10mm}
\centerline{ \desepsf(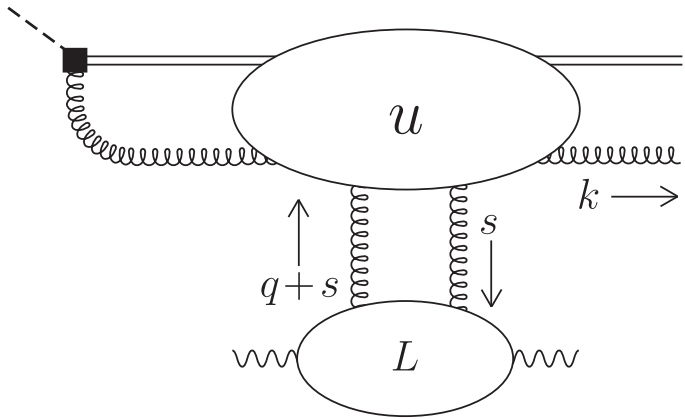 width 9 cm) }
\vspace{10mm}
\caption{Structure of the diffractive gluon distribution. 
The evolution of the parton system and the interaction with the 
hadron's field are evaluated in the lowest perturbative 
approximation. }
\label{fig:ajm}
\end{figure}

\vspace{10mm}

In general,  aligned jet configurations may involve  
$N$ final state partons, with the hadron's field being represented by 
an external color  field. 
The problem 
is then that of  calculating 
an amplitude of the type   
\begin{equation}
{\cal M}^{j {s_1} {\cdots} {s_N}} = \int dy^- e^{i\beta x_\pom p_A^+ y^-}
\langle k_1,s_1,\dots,k_N,s_N | 
{\cal O}^{j}(0,y^-,{\bf 0})|0\rangle_{\cal A} \hspace*{0.2 cm} ,   
\label{Mgeneralized}
\end{equation}
where ${\cal O}^{j}$ represents the measurement 
operator (defined by the gluon or quark distribution),  
$k_i,s_i$ are the momenta and spins of  final state
partons  
and the matrix element is evaluated 
 in the presence of an external color field $\cal A$. 
As $1 / x_\pom \to \infty$, 
the diffractive parton distribution is obtained~\cite{hkslong} by 
integrating over the final states 
the squared amplitude, taken  
in 
 the limit in which 
the momenta $k_i^-$ become 
large (scaling like $1 / x_\pom$), 
while  the external field $\cal A$ stays fixed.

At high energy the amplitude ${\cal M}$ may be 
evaluated  using the approximation in which the action of the external 
field  is simply to produce an eikonal phase $F$ 
for each parton while leaving 
its minus momentum and its transverse position unchanged: 
\begin{equation}
F({\bf b}) \equiv
{\cal P}\ \exp\left\{
-ig\int_{-\infty}^{+\infty} d z^- {\cal A}_a^+(0,z^-,{\bf b})\, t_a
\right\} \hspace*{0.2 cm} .
\label{Fdef}
\end{equation}
(For related approaches, see refs.~\cite{buch} and~\cite{bartetal}.) 
Then  the answer for ${\cal M}$ decomposes into a part associated with 
$F$ and a part associated with 
the measurement operator~\cite{hkslong}.  
The result for $N = 1$  
has the form 
\begin{equation}
{\cal M}^{js} =
-\int {d^2 {\bf p}\over (2\pi)^2}\
\left[
\widetilde F({\bf k}-{\bf p})\
F(0) - (2\pi)^2 \, \delta^2 ( {\bf k}-{\bf p}) 
\right]
\psi^{js}({\bf k},{\bf p}) \hspace*{0.2 cm} , 
\label{M0}
\end{equation}
where $\widetilde F$ is the  Fourier transform of $F$, 
\begin{equation}
\widetilde F({\bf q} ) =
\int d^2{\bf b}\
e^{i {\bf q} \cdot {\bf b}}
F({\bf b}) \hspace*{0.2 cm}  ,   
\label{Ftilde}
\end{equation}
and $\psi$ is the light-cone wave function 
\begin{equation}
\label{wavefctndef}
\psi^{js}({\bf k},{\bf p}) = 
i\ { \langle k^-,{\bf p};s|
{\cal O}^j(0)
|0\rangle 
\over \beta x_\pom p_A^+ + {\bf p}^2/(2k^-)} \hspace*{0.3 cm} .  
\end{equation}
This wave function can be interpreted 
as representing 
 the parton state just before it interacts with the 
external field. The explicit expressions for the gluon and quark 
wave functions may be found in refs.~\cite{hkslong,hkslett}. 
It is worth remarking that these wave functions are associated 
with the operator in Eq.~(\ref{opg}) (or its analogue for quarks) 
rather than with the electromagnetic current of the 
deeply inelastic scattering. In particular, they do not depend 
on $Q^2$. The dependence on $Q^2$ is factored out in the coefficients 
$\hat F$ in Eq.~(\ref{fact}). This allows one to systematically 
include perturbative corrections beyond the leading logarithms 
for both the quark and the gluon contributions. 

A technical aspect of the light-cone 
formalism is also worth noting.    
 The wave function $\psi$ describes a system 
with large minus momentum. The natural  formulation 
for such a system is 
one in which the theory is quantized 
on planes of equal $x^-$. This singles out the  components 
of the quark field that give the independent degrees of freedom.  
It is  an equation of constraint~\cite{alignedjet}  
that allows one to  
relate these to the components  
that have a simple partonic interpretation 
for a hadron  moving in the plus direction.

\section{The case of small-size states: perturbation expansion }

 In Eq.~(\ref{M0})  the  interaction of the parton system 
created by the measurement operator with the hadron's color field is 
described by the factor in the square brackets.  
If one considers a thought experiment on  
a color source with sufficiently small radius,  
 this interaction can be represented as   
a perturbation expansion. Then the full answer for the 
diffractive distribution can be computed at weak coupling. 
The result for the 
example of a small color dipole made of a pair of heavy quarks 
 may be found in ref.~\cite{hkslett}. The form of the 
result at the lowest 
perturbative order is very simple.   It depends on the size of 
the source, $ R \sim 1/M$, through an overall quadratic factor 
$1 / M^2$.  It  is  also scale invariant, since all graphs are 
ultraviolet convergent.

To higher loops, ultraviolet divergences 
arise.  The 
renormalization of these divergences 
brings in a renormalization scale $\mu$ 
and  gives rise  to scaling violation. 
The dependence of the diffractive parton distributions on   
 $\mu$ can be included systematically 
through the usual DGLAP evolution equations. 

The higher loop contributions are 
of order $\alpha_s \ln (\mu^2 / M^2) $ compared to the lowest order 
term. When $\ln (\mu^2 / M^2) $ is large, they are important,  
and thus evolution is important. When $\mu \approx M$, they are small 
corrections to the lowest order term. Then the 
diffractive parton distributions at a scale $\mu$ are obtained 
by solving the  evolution equations with the 
results of the lowest order calculation as 
a boundary condition at a scale of order $M$.

\section{The realistic case:  semihard dominance and evolution  }

In  the case of realistic 
experiments of diffraction 
on large size systems,  
 the interaction with the hadron's field depends on 
nonperturbative physics. 
The answer for the diffractive parton distributions 
cannot be determined from a purely perturbative expansion, but it 
depends on the infrared behavior of the diffraction process. 

 Suppose one started with the 
small size situation of the previous section and let the 
system's size increase.  
A possible scenario 
could be that, as the system's size becomes of order 
$R \approx 1 / ( 300 \, {\mbox {MeV}})$, the answer becomes 
completely dominated by the soft region   
$ k_\perp \sim 300 \, {\mbox {MeV}}$. This is what a    
perturbative power counting  
 for the diffractive scattering amplitude would suggest. 
This would suggest  that 
the diffractive parton distributions in the case of 
diffraction on protons are radically 
different from those for a small system.

A different scenario is discussed in ref.~\cite{hkslong}.  
In this scenario,  an 
intermediate scale  exists 
at which nonperturbative dynamics sets in that reduces  
the infrared sensitivity 
suggested by the perturbative 
power counting. 
Then 
the distance 
scales that dominate the diffraction process, rather than 
continuing to grow as we go to larger and larger sizes,  
stay of the order of some  semihard scale $ M_{\rm SH}^{-1}$   
of the order of a GeV$^{-1}$. 
Under this hypothesis, the diffractive parton 
distributions should be similar to what one gets 
from the distributions 
 for a small system   using evolution from the scale $ M = 
 M_{\rm SH}$ to the multi-GeV scale relevant for experiments.

  Results   for the  $\beta$ and $Q^2$ behavior 
of diffractive deeply inelastic scattering 
are shown in Fig.~2 and compared with the 
ZEUS data of ref.~\cite{zeusf2d2}. 
 The curves in this figure are 
obtained for  $ M_{\rm SH} = 1.5 \, {\mbox {GeV}}$. 
 Given the value of $M$, both  
 the $ Q^2 $ dependence 
 and also the $\beta$ dependence are determined from theory. 
Only the overall normalization  is not. 
The agreement between theory and experiment in Fig.~2 is 
evidently not perfect; however, certain qualitative features 
of the data are reproduced. 
The curves of Fig.~2 show that 
in the mid range of $\beta$ 
the diffractive structure function $F_2^{\rm diff}$ is rather flat as 
a function of $\beta$ 
and, in the range of $Q^2$ shown, it 
increases with 
 $Q^2$ for $\beta $ up to about  0.5. 
These results are dramatically at variance with what happens in the 
case of the inclusive $F_2$. 
The difference in the pattern of scaling violation between the 
diffractive case and the inclusive case is illustrated in 
Fig.~3. 

\begin{figure}[htb]
\vspace{100mm}
\includegraphics{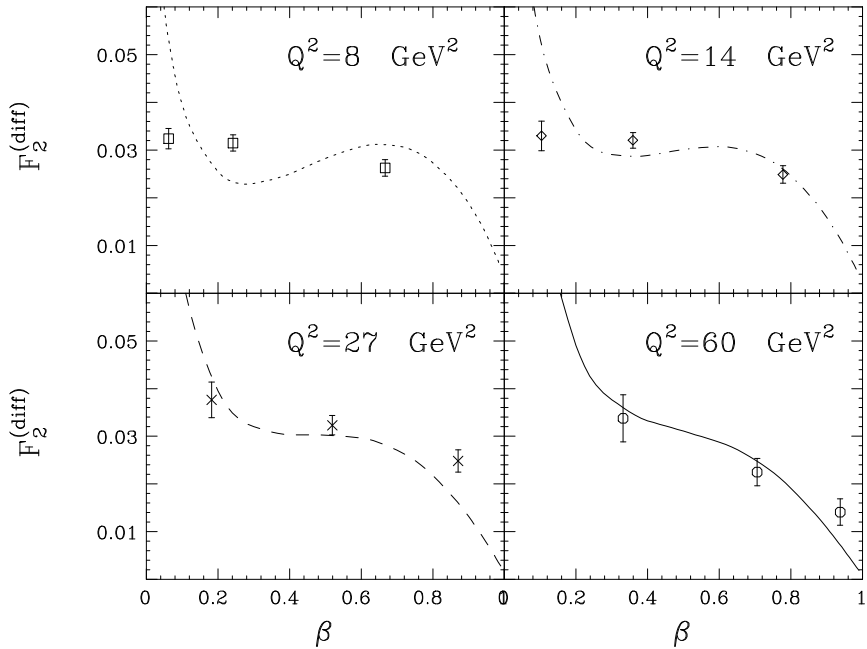}
\caption{ The $\beta$ dependence of the diffractive structure function 
$F_2^{\rm diff}$ for different values of $Q^2$. 
Also shown are the ZEUS data. }
\label{fig:betadep}
\end{figure}

The reason for these behaviors lies with the form (\ref{M0}) 
of the diffractive parton distributions. The explicit 
evaluation~\cite{hkslong}  of this form shows that   
the nature of the coupling of the hadron's field to fast moving 
 gluons or quarks is such that the $\beta$ dependence is 
rather mild. Also, it shows that 
the hadron's field has a much stronger coupling 
to gluons than it does to quarks,  with the 
ratio being  controlled in the first approximation by the ratio of 
color factors 
$ C_A^2 \ (N_c^2 - 1) / (C_F^2 \ N_c) = 27 / 2$~\cite{hkslett}.   
Thus the depletion of quarks in the mid range of momentum fractions 
under evolution, due to gluon bremsstrahlung, is compensated by the 
replenishment of quarks from gluon splitting 
into $q \bar q$ pairs. 
The net result is that the diffractive quark distribution 
 rises with $Q^2$ at intermediate $\beta$, and so does 
 $F_2^{\rm diff}$.

\begin{figure}[htb]
\vspace{100mm}
\includegraphics{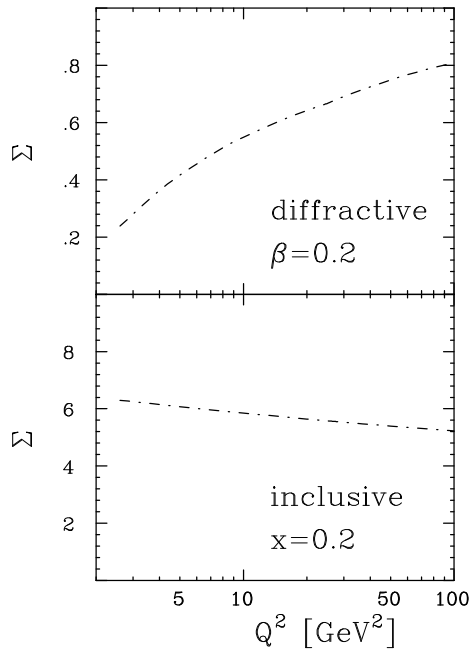}
\caption{ The $Q^2$ dependence of the flavor singlet quark 
distribution 
$\Sigma$ at moderate values of momentum fractions.  }
\label{fig:sv}
\end{figure}

The predictions of Fig.~2 are  based on the choice 
$ M =  1.5 \, {\mbox {GeV}}$. 
If one decreases this value below $1 \, {\mbox {GeV}}$, 
one finds~\cite{hkslong} that, 
compared to the data,  
there is too much of a slope in $\beta$ and too strong a 
dependence on $Q^2$. 
Thus a semihard scale seems to be preferred by the data. 
   Note that a semihard scale in 
 diffractive deeply inelastic scattering 
 is also consistent with experimental 
observations 
of the $x_\pom$ dependence.   
The value of 
$\alpha_\pom (0) - 1$   
measured in diffractive deeply inelastic 
scattering~\cite{zeusf2d2,h1scalviol}  differs by a factor of $2$ 
from the corresponding value measured in soft hadron-hadron cross 
sections~\cite{muedis98}.

The origin of the scale $M_{\rm SH}$ is essentially 
nonperturbative. This scale is generated in a region of 
low $x$ and low $Q^2$ where the gluon density is high 
and one may expect nonlinear effects to set in~\cite{ffss}.   
It would be interesting to study whether such a  scale arises 
in other contexts 
 associated with the growth of the gluon density.

\section*{Acknowledgments}
We  thank the organizers for their invitation to 
ISMD99. This research is supported in part by the US Department 
of Energy.

\end{document}